\documentclass{cernepprep} 
\usepackage{cernunits,cernchemsym,heppennames2}
\usepackage{cite,fancyvrb}

\usepackage{graphicx}
\usepackage{dcolumn}
\usepackage{bm}
\usepackage[utf8]{inputenc}
\usepackage[T1]{fontenc}
\usepackage{newtx}
\usepackage{etoolbox}
\usepackage{amsmath}
\usepackage{import}
\usepackage{enumitem}
\usepackage{subfigure}
\usepackage[subfigure]{tocloft}
\usepackage{multirow}
\usepackage{overpic}
\usepackage{tabularx}
\usepackage{setspace}
\usepackage{anyfontsize}
\usepackage[T1]{fontenc}
\usepackage{gensymb}
\usepackage{hyperref}

\begin{document}

\begin{titlepage}
\EPnumber{2025-195}
\EPdate{\today}
\DEFCOL{CDS-Library}

\title{Measured Properties of an Antihydrogen Beam}

E.~D.~Hunter$^{1,2}$\footnote{Electronic mail: eric.david.hunter@cern.ch},
M.~Bumbar$^{1,2,3}$,
C.~Amsler$^{4}$,
M.~N.~Bayo$^{5,6}$,
H.~Breuker$^{7}$, 
M. Cerwenka$^{4,3}$,
G. Costantini$^{8,9}$,
R. Ferragut$^{5,6}$,
M. Giammarchi$^{4}$,
A. Gligorova$^{4}$\footnote{Current Address: Faculty of Physics, University of Vienna, Boltzmanngasse 5, 1090 Vienna, Austria}, 
G. Gosta$^{8,9}$,
M.~Hori$^{2,10}$,
C.~Killian$^{4}$,
V.~Kraxberger$^{4,3}$,
N.~Kuroda$^{11}$,
A.~Lanz$^{4,3}$\footnote{Current Address: University College London, London WC1E 6BT, United Kingdom}, 
M.~Leali$^{8,9}$,
G.~Maero$^{12,6}$,
C.~Mal\-bru\-not$^{1}$\footnote{Current Address: TRIUMF, Vancouver BC V6T 2A3, Canada},
V.~Mascagna$^{8,9}$,
Y.~Matsuda$^{11}$,
S.~Migliorati$^{8,9}$,
D.~J.~Murtagh$^{4}$,
M.~Romé$^{12,6}$,
R.~E.~Sheldon$^{4}$,
M.~C.~Simon$^{4}$,
M.~Tajima$^{13,2}$,
V. Toso$^{6,12}$,
S.~Ulmer$^{7,14}$,
L.~Venturel\-li$^{8,9}$,
A.~Weiser$^{4,3}$,
E.~Wid\-mann$^{4}$\\

\centering{(The ASACUSA-Cusp Collaboration)}\\[20pt]

$^1$CERN, 1211 Geneva 23, Switzerland,
$^2$Imperial College London, London SW7 2BW, United Kingdom,
$^3$Vienna Doctoral School in Physics, University of Vienna, 1090 Vienna, Austria,
$^4$Stefan Meyer Institute for Subatomic Physics, Austrian Academy of Sciences, 1010 Vienna, Austria,
$^5$L-NESS and Department of Physics, Politecnico di Milano, 22100 Como, Italy,
$^6$INFN sez. Milano, 20133 Milan, Italy,
$^7$Ulmer Fundamental Symmetries Laboratory, RIKEN, 351-0198 Saitama, Japan,
$^8$Diparti\-mento di Ingegneria dell'In\-formazione, Universit\`a degli Studi di Brescia, 25123 Brescia, Italy,
$^9$INFN sez. Pavia, 27100 Pavia, Italy,
$^{10}$Max-Planck-Insitut f\"ur Quantenoptik, D85748 Garching, Germany,
$^{11}$Institute of Physics, Graduate School of Arts and Sciences, University of Tokyo, 153-8902 Tokyo, Japan
$^{12}$Dipartimento di Fisica, Università degli Studi di Milano, 20133 Milan, Italy
$^{13}$Japan Synchrotron Radiation Research Institute, 1-1-1 Kouto, Sayo-cho, Sayo-gun, Hyogo 679-5198, Japan
$^{14}$Insitut f\"ur Experimentalphysik, Heinrich Heine Universit\"at, D\"usseldorf, Germany

\date{\today}

\begin{abstract}
We report a factor of $100$ increase in the antihydrogen beam intensity downstream of ASACUSA's Cusp trap: 320 atoms detected per 15-minute run. The beam contains many Rydberg atoms, which we selectively ionize to determine their velocity and binding energy. The time of flight signal is modeled using a 1D Maxwellian velocity distribution with a temperature of $1500\,\mathrm{K}$, which is close to the measured antiproton plasma temperature. A numerical simulation reproduces the observed distribution of binding energies and suggests that about $16\%$ of the atoms may be in the ground state.
\end{abstract}

\end{titlepage}

Within the Standard Model of particle physics, the combined operation of charge conjugation (matter $\leftrightarrow$ antimatter), parity inversion ($x,y,z \rightarrow -x,-y,-z$), and time reversal ($t\rightarrow -t$) known as CPT is an exact symmetry. One consequence of CPT symmetry is that atomic transition frequencies in hydrogen (H) and antihydrogen ($\overline{\mathrm{H}}$) should be identical. For example, the ground state hyperfine interval (${}^3\mathrm{S}_1 \rightarrow {}^1\mathrm{S}_0$) of $\overline{\mathrm{H}}$ matches that of H to a precision of $10^{-4}$ \cite{ahmadi_2017_observation}. The interval is known at the $10^{-12}$ level in H \cite{karshenboim_2005_precision}. Members of our collaboration recently measured this quantity to $4\times 10^{-10}$ in a beam of H \cite{nowak_2024_cpt}, and we are preparing to measure it with a beam of $\overline{\mathrm{H}}$ in an equivalent spectroscopy apparatus \cite{widmann_2004_measurement,widmann_2013_measurement}. In the present article, we characterize the velocity and binding energy of atoms in an $\overline{\mathrm{H}}$ beam and evaluate their usefulness for the anticipated hyperfine measurement. Compared to previous studies of the velocity \cite{gabrielse_2004_first} and binding energy \cite{gabrielse_2002_driven,kolbinger_2021_measurement} of nascent $\overline{\mathrm{H}}$ atoms, our beam is over $100$ times more intense, allowing a time of flight spectrum and continuous binding energy distribution to be acquired for the more easily ionized states.

To create $\overline{\mathrm{H}}$, the most efficient scheme so far is ``slow-merge mixing'' as used by the ALPHA Collaboration \cite{ahmadi_2017_antihydrogen}. Antiprotons ($\overline{\mathrm{p}}$) from CERN's ELENA facility \cite{maury_2014_elena} and positrons ($\mathrm{e^+}$) from a radioactive source \cite{lanz_2023_upgrade} are confined in two neighboring electrodes of a Penning-Malmberg trap \cite{gabrielse_1989_possible}. Initially, the particles are prevented from interacting by an electrostatic barrier between them. When the $\overline{\mathrm{p}}$ are slowly raised over this barrier, they enter the $\mathrm{e^+}$ plasma and produce $\overline{\mathrm{H}}$, mainly via 3-body recombination: two $\mathrm{e^+}$ collide close to one $\overline{\mathrm{p}}$, one $\mathrm{e^+}$ is bound to the $\overline{\mathrm{p}}$, and the other $\mathrm{e^+}$ carries away the binding energy (and any excess kinetic energy) \cite{robicheaux_2008_atomic}. 

This process produces atoms in excited states with quantum number $n\gg 1$. The bound $\mathrm{e^+}$ can lose more energy in subsequent collisions between $\overline{\mathrm{H}}$ and other $\mathrm{e^+}$ in the plasma \cite{robicheaux_2004_simulations}. These deexcitation collisions, as well as the 3-body reaction, occur more frequently in colder $\mathrm{e^+}$ plasma. The results presented here rely on improved cooling methods for large numbers of $\mathrm{e^+}$ in ASACUSA's Cusp trap \cite{amsler_2022_reducing,hunter_2025_best}.

We produce about $2\times 10^6$ $\overline{\mathrm{H}}$ every $15$ minutes. This estimate is based on the number of $\overline{\mathrm{p}}$ used and the fraction of annihilations that occur during mixing \cite{hunter_2025_best}. About $320$ of these atoms are detected after exiting the Cusp trap in a beam of fractional solid angle of $2.1\times 10^{-4}$, which is within the acceptance of our spectroscopy line \cite{diermaier_2017_inbeam}. We use field ionization \cite{vrinceanu_2004_strongly} to analyze the beam velocity $\upsilon$ and binding energy, indexed by $n$. From the shape of the $\upsilon$ and $n$ distributions, we estimate the number of atoms with $\upsilon<1500\,\mathrm{m/s}$ and the ground state ($n=1$) fraction.  We also observe that lower values of $n$ appear to be correlated with lower $\upsilon$, which is plausible as slower atoms spend more time colliding with $\mathrm{e^+}$ inside the plasma. 



\begin{figure}
    \centering
    \includegraphics[width=0.95\linewidth]{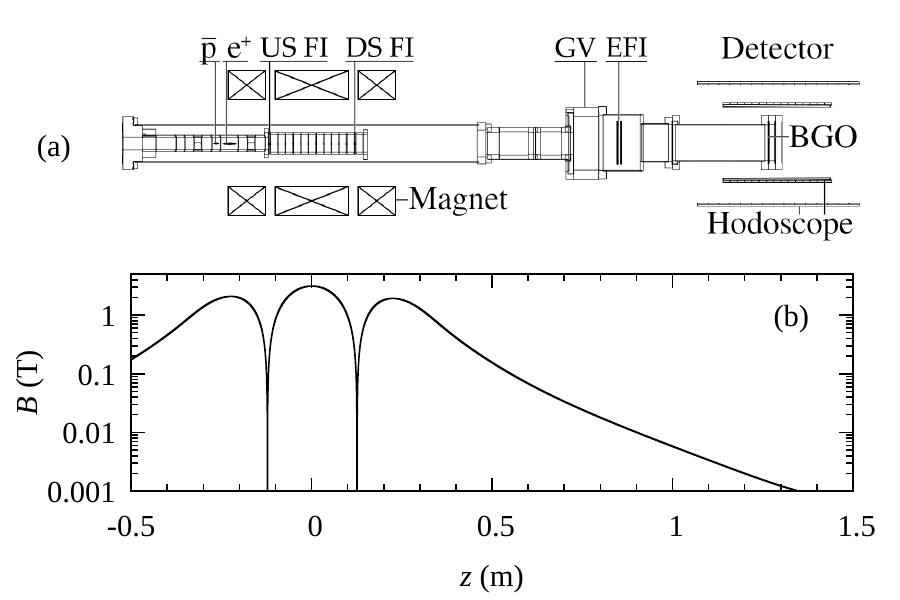}
    \caption{Simplified cross section of the experiment (a) and calculated magnetic field $B$ along the axis of symmetry (b). Scale drawing in (a) is sized to match the axial positions $z$ in (b). The downstream direction corresponds to increasing $z$. Abbreviations: upstream (US) and downstream (DS) field ionizer (FI), gate valve (GV), external FI (EFI), bismuth germanium oxide crystal (BGO), scintillating bars (Hodoscope), Cusp magnet coils (Magnet). We use the symbols $\overline{\mathrm{p}}$ and $\mathrm{e^+}$ for the plasmas.}
    \label{fig:appa}
\end{figure}

Figure~\ref{fig:appa} shows a wireframe drawing of the Cusp trap and analysis beamline. The double anti-Helmholtz coils of the Cusp magnet produce two nulls in the magnetic field: $B=0$ at axial position $z=-\,0.11$ and $+\,0.11\,\mathrm{m}$. The strong $B$ gradients focus low-field-seekers\textemdash atoms with magnetic moment vector opposite to $\vec{B}$\textemdash and defocus high-field-seekers \cite{nagata_2014_a,nagata_2015_the}. The trap electrodes have inner diameter $34\,\mathrm{mm}$. The entrance and exit to the trap are screened by $79\%$ transparent copper meshes to protect the $\mathrm{e^+}$ inside the $6\,\mathrm{K}$ trap from external, room-temperature radiation at the cyclotron frequency \cite{amsler_2022_reducing}. The antiproton lifetime exceeds $6000\,\mathrm{s}$, implying a vacuum below $10^{-13}\,\mathrm{mbar}$ \cite{sellner_2017_improved}. 

$\overline{\mathrm{H}}$ is formed inside a $\mathrm{e^+}$ plasma located at the $B=2\,\mathrm{T}$ maximum upstream of the first field null. We first transfer $1.2\times 10^8\,\mathrm{e^+}$ and $2.5\,\times 10^6$ $\overline{\mathrm{p}}$ to the Cusp trap from separate accumulation traps. The $\mathrm{e^+}$ plasma is about five times colder than in previous experiments by ASACUSA \cite{amsler_2022_reducing} and the number of $\overline{\mathrm{p}}$ is at least four times greater \cite{kolbinger_2021_measurement}. The plasmas are compressed, cooled, and purified to remove ions from the $\mathrm{e^+}$ and electrons from the $\overline{\mathrm{p}}$. These operations take about ten minutes. Then the $\overline{\mathrm{p}}$ are combined over $60\,\mathrm{s}$ with the $\mathrm{e^+}$. A long ``mixing'' time seems necessary because of the large amount of antimatter in our trap: $50$ times more $\mathrm{e^+}$ and $25$ times more $\overline{\mathrm{p}}$ compared to ALPHA's experiments, where the plasmas are merged in $1\,\mathrm{s}$ \cite{ahmadi_2017_antihydrogen,baker_2021_laser,baker_2025_precision}. The plasma effects that favor slower mixing are beyond the scope of the present article and will be explored elsewhere. More information about ASACUSA's traps, and the steps for catching, transferring, and tailoring the $\overline{\mathrm{p}}$ and $\mathrm{e^+}$ plasma, is given in Refs.~\cite{amsler_2022_reducing,hunter_2025_best,lanz_2023_upgrade,hunter_2023_sdr}.

$\overline{\mathrm{H}}$ is detected if it escapes the plasma downstream into a cone of half-angle $1.7\degree$, assuming that it follows a straight-line path (low-field-seekers at half-angle up to $6\degree$ may reach the detector because of the Cusp magnet's focusing effect \cite{nagata_2014_a}). The detector uses a 3-coincidence counter requiring one hit in a bismuth germanium oxide disk (BGO) of radius $4.5\,\mathrm{cm}$ and thickness $0.5\,\mathrm{cm}$ \cite{nagata_2017_development} and one hit in each $0.5\,\mathrm{cm}$ thick layer of scintillating bars, located $11$ and $17\,\mathrm{cm}$ from the axis of symmetry (Hodoscope) \cite{kolbinger_2018_recent}. For the BGO, a hit is at least $1\,\mathrm{MeV}$ of deposited energy detected by any of the 128 silicon photomultipliers (AFBR-S4N44C013) distributed over the crystal. For the bar layers, a hit is a coincidence between silicon photomultipliers at both ends of any of the 32 bars in that layer \cite{kraxberger_2023_upgrade}. The energy spectrum for $\overline{\mathrm{H}}$ hitting the BGO matches the one we obtain by sending bare $\overline{\mathrm{p}}$ downstream with enough kinetic energy to reach the detector ($50$ to $100\,\mathrm{eV}$). If we repeat these tests with lower-energy $\overline{\mathrm{p}}$ ($1$ to $3\,\mathrm{eV}$), or with the gate valve (GV) closed, we get a steep decreasing spectrum with mean energy five times lower. See Appendix~\ref{sec:appx_hodo}. The GV-closed signal is used to determine the background rate during mixing for the 3-coincidence counter, which can be triggered by cosmic rays or by pions from $\overline{\mathrm{H}}$ and $\overline{\mathrm{p}}$ annihilating in the trap.  

Two sets of field ionizers (FI), the upstream FI and downstream FI at $z=-\,0.12$ and $+\,0.12\,\mathrm{m}$, can produce an electric field $F$ up to $7\times 10^{-7}$ in atomic units (roughly $350\,\mathrm{V/mm}$, averaged in the transverse plane). The external FI (EFI), described in Ref.~\cite{kolbinger_2021_measurement} and used here only for validation, is located at $z=0.9\,\mathrm{m}$, where $B\approx0.01\,\mathrm{T}$. The in-trap FI enclose regions with higher values of $B$, which makes high-$n$ states harder to ionize \cite{jonsell_2009_simulation}. We use a classical orbit simulation to find the $F$ needed to ionize atoms of given binding energy for all possible values of the orbital ellipticity and the direction of $F$. This produces a range of $F$ for a given $n$. Our results agree with the $B=0$ expression (in atomic units) \cite{rakovic_1998_ionization}
\begin{equation}
    \frac{1}{7.7 n^4} < F < \frac{1}{2.6 n^4}
    \label{eq:n}
\end{equation}
for $n < 40$ and $B<0.35\,\mathrm{T}$. $B$ at the upstream and downstream FI depends on the radius and is in the range $0.1<B<0.35\,\mathrm{T}$. The FI geometry, simulation, and validation, including comparisons between $\overline{\mathrm{H}}$ ionization measurements using upstream, downstream, and external FI, are given in Appendix~\ref{sec:appx_fi}.



The distance from the upstream FI to the detector is $1.41\,\mathrm{m}$. By varying the voltage on the FI, we can chop the beam and measure the time of flight (ToF) for the subset of ionizable $\overline{\mathrm{H}}$. $F$ is a function of space and the upstream FI voltage $V_\mathrm{US}$, and each value of $F$ comes with a distribution of $n$ that may be ionized, as in Eq.~(\ref{eq:n}). We include both effects in our analysis, but in the text below we simply give $90\%$ average ionization thresholds for $n$ (Appendix~\ref{sec:appx_fi}). We study the distribution of $n$ and $\upsilon$ by periodically pulsing or ramping the upstream FI and blocking (or not) with the downstream FI, according to the three protocols shown in Fig.~\ref{fig:beam}:

\begin{figure}
    \centering
    \includegraphics[width=0.94\linewidth]{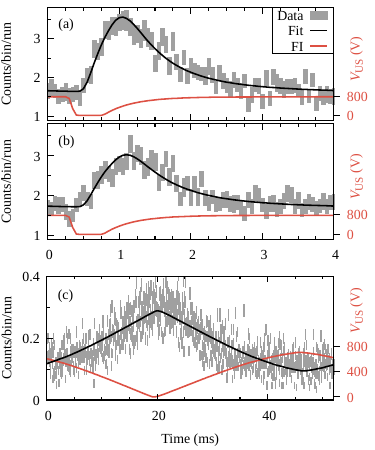}
    \caption{Detector counts vs.\ time for three periodic ionization protocols: (a) pulsed without blocking, (b) pulsed with blocking, (c) triangle wave. Detector counts (``Data'') are summed for each period and averaged over about 100 mixing runs, with bins centered at their means and extending vertically by one standard deviation of the mean in both directions. A constant background of $80\,\mathrm{counts/run}$ is observed during mixing runs with the GV closed. This number is divided by the number of bins in each trace and subtracted from each $50\,\mathrm{\mu s}$ bin.}
    \label{fig:beam}
\end{figure}

\begin{enumerate}[label=(\alph*),leftmargin=*,align=left]
    \item Pulsed without blocking: $V_\mathrm{US}=800\,\mathrm{V}$, pulsed to $0\,\mathrm{V}$ for $400\,\mathrm{\mu s}$ at a repetition rate of $240\,\mathrm{Hz}$. The component of the beam with $n > 27$ is chopped by this protocol. Field ionized atoms annihilate close to the FI, producing no signal at the detector.
    \item Pulsed with blocking: $V_\mathrm{US}$ is pulsed as before and the downstream FI voltage is held fixed at $135\,\mathrm{V}$ so that only atoms with $27<n<36$ contribute to the time-dependent part of the ToF signal. 
    \item Triangle wave: $V_\mathrm{US}$ is ramped linearly between $0$ and $800\,\mathrm{V}$ at $19\,\mathrm{Hz}$. This is used to find the beam intensity as a function of $V_\mathrm{US}$, after accounting for the delay due to the ToF between FI and detector. If we fit the signal to a linear function of $V_\mathrm{US}$ with variable delay, we find that the residual sum of squares is minimized for a delay of $0.65\pm 0.05\,\mathrm{ms}$.
\end{enumerate}

We model the ToF signal in (a,b) by using (c) to find the fraction of signal atoms that pass the FI as $V_\mathrm{US}$ varies in time and adding the time from the FI to the detector as $1.41\,\mathrm{m}/\upsilon$ (Appendix~\ref{sec:appx_tof}). We assume that $\upsilon$ is distributed as $\mathrm{exp}[-\upsilon^2/2\upsilon_t^2]$, a 1D Maxwellian with thermal velocity $\upsilon_t=\sqrt{k_\mathrm{B}T_\mathrm{m}/m}$, where $T_\mathrm{m}$ minimizes the residual sum of squares, producing the fits shown in the figure. We expect a 1D Maxwellian because $\overline{\mathrm{p}}$ are sampled from the $\overline{\mathrm{p}}$ plasma when they have enough axial energy to pass over the electrostatic barrier into the $\mathrm{e}^+$. The transverse velocity components integrate out \cite{eggleston_1992_parallel} and they are not relevant for selecting a beam-like component because they thermalize quickly once the $\overline{\mathrm{p}}$ enters the $\mathrm{e}^+$ plasma \cite{hurt_2008_positron}. 
 
The fitted $T_\mathrm{m}$ are $1510\pm190\,\mathrm{K}$ for (a) and $1150\pm140\,\mathrm{K}$ for (b). These values are compatible with the measured $\overline{\mathrm{p}}$ plasma temperature, which is $1200\pm 500\,\mathrm{K}$ (mean and standard deviation) during mixing. The uncertainties on $T_\mathrm{m}$ are given as the standard deviation for 10000 bootstrap samples in each case, sampling with replacement from 85 runs for (a) and 103 runs for (b) \cite{efron_1979_bootstrap}. 

From the $\upsilon$ distribution we derive the fraction of atoms with $\upsilon<1500\,\mathrm{m/s}$: $33.2\pm 1.8\%$ for (a) and $37.5\pm 2.1\%$ for (b), implying that $1/3$ of the atoms are slow enough to be used in the ASACUSA spectrometer \cite{diermaier_2017_inbeam}. The average ${<}\upsilon{>}$ is $2770\pm 160\,\mathrm{m/s}$ for (a) and $2440\pm 150\,\mathrm{m/s}$ for (b). Thus, low $n$ is weakly correlated with low $\upsilon$. Slower atoms take longer to leave the plasma, so they have more time to decay to lower $n$ via collisions with the $\mathrm{e}^+$. We can also estimate ${<}\upsilon{>}$ by dividing the distance from the FI to the detector by the delay fitted in (c). We get $1.41\,\mathrm{m} / 0.65\,\mathrm{ms}=2170\,\mathrm{m/s}$, which is lower than the ToF analysis results. This is additional evidence that lower $n$ is correlated with lower $\upsilon$, since the triangle wave blocks low-$n$ states for less time than it blocks high-$n$ states.

The triangle wave protocol analyzes the binding energy of atoms with $n>27$, but it does not give the number of atoms with $n=1$, which is needed for spectroscopy. We predict that number using a numerical simulation \cite{radics_2014_scaling}, which takes as input (i) the $\mathrm{e^+}$ plasma temperature and density, (ii) $B$, (iii) the amount of time for a $\overline{\mathrm{p}}$ to pass through the plasma, (iv) the blackbody radiation temperature, and (v) an upper limit $n_\mathrm{lim}$, which affects the overall normalization. The quantities (i), (ii), and (iii) are constrained by our measurements. The $\mathrm{e^+}$ temperature is $44\pm 5\,\mathrm{K}$ and $B=2.0\pm 0.1\,\mathrm{T}$ \cite{hunter_2025_best}. The density $1.8\times 10^8\,\mathrm{cm^{-3}}$ is calculated using known properties of the plasma and the trap \cite{christensen_2024_exploiting}. For the transit time we use the $\mathrm{e^+}$ plasma length divided by $\upsilon$, which is taken from the same 1D Maxwellian as in Fig.~\ref{fig:beam}(a). We will see that the ground state fraction varies by about $1\%$ for sensible choices of (iv) and (v). Thus, our prediction for $n=1$ does not depend on any free parameters.

\begin{figure}
    \centering
    \includegraphics[width=0.95\linewidth]{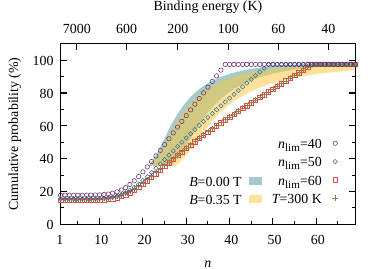}
    \caption{Simulated cumulative probability distribution for $n$ and comparison with FI measurements (yellow band). Four conditions are shown: $n_\mathrm{lim}=40,50,60$ for $15\,\mathrm{K}$ blackbody radiation, and $n_\mathrm{lim}=60$ for $300\,\mathrm{K}$. Data from Fig.~\ref{fig:beam}(c) is plotted with $n$ ranges found in Appendix~\ref{sec:appx_fi} for $B=0$ and $B=0.35\,\mathrm{T}$.}
    \label{fig:ndist}
\end{figure}

The simulation gives the probability for one $\overline{\mathrm{p}}$ to become part of one $\overline{\mathrm{H}}$ of a given $n$ after a single pass through the $\mathrm{e^+}$ plasma. Although $\overline{\mathrm{p}}$ may pass thousands of times before forming stable $\overline{\mathrm{H}}$, only the final pass determines the properties of the atom that escapes. The probability is normalized to its sum for all $n$ values up to $n_\mathrm{lim}$. In Fig.~\ref{fig:ndist} we plot the cumulative probability (sum from $1$ to $n$) for three different values of $n_\mathrm{lim}$ and for blackbody temperature $15$ and $300\,\mathrm{K}$. Averaging the predictions from the four models shown in Fig.~\ref{fig:ndist}, we find that for all $\overline{\mathrm{H}}$ formed, $16\pm2\%$ should be in $n=1$.

We compare the simulation result with Fig.~\ref{fig:beam}(c), fitting a spline to the data points and converting $V_\mathrm{US}$ to $n$ ranges, for two limiting $B$ values, as described in Appendix~\ref{sec:appx_fi}. Simulation and measurement agree that $20<n<50$ for most of the atoms. The agreement seems best for $n_\mathrm{lim}\approx 45$. This value seems too low: atoms with $n\lesssim80$ should survive the stray electric fields between the plasma and the detector. One possible explanation is that the $B$ gradients in the Cusp remove atoms with a large magnetic moment from the beam. The average absolute value of the magnetic moment increases with $n$ in a model-dependent way \cite{robicheaux_2006_threebody} and also changes along the trajectory \cite{gallagher_1994_rydberg,lundmark_2015_towards}, so we do not explicitly include this effect. 

The $\overline{\mathrm{p}}$ beam intensity delivered by ELENA is steadily increasing \cite{bojtar_2024_jacow}. To find out how the $\overline{\mathrm{H}}$ beam rate scales with the number of $\overline{\mathrm{p}}$, we insert a variable number of secondary electron multiplier (SEM) grids in the ELENA beamline, reducing the intensity by about $9\%$ per grid \cite{hori_2005_photocathode}. We further extend the range by reducing the number of $\overline{\mathrm{p}}$ bunches caught from ELENA from $3$ to $1$. We alternate $\overline{\mathrm{H}}$ production runs with runs where we count the number of $\overline{\mathrm{p}}$ by slowly extracting them upstream to a microchannel plate, phosphor screen, and silicon photomultiplier detection system \cite{hunter_2020_plasma,cripe_2024_summer}. We take about $9$ runs of each kind for $9$ different experimental conditions (number of SEM grids, number of bunches). 

\begin{figure}
    \centering
    \includegraphics[width=0.92\linewidth]{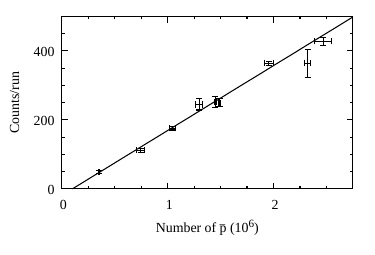}
    \caption{Total number of detector counts per mixing run as a function of the number of $\overline{\mathrm{p}}$. Points and error bars are the mean and the standard deviation of the mean for data taken in 9 different experimental conditions. The background rate (with GV closed) is not subtracted because it is only known for the highest number of $\overline{\mathrm{p}}$. The trendline is fit using York's algorithm \cite{york_1966_leastsquares,vermeesch_2018_isoplotr}, which correctly accounts for the simultaneous uncertainty in both coordinates. The slope is $188\pm 5$ counts per $10^6$ $\overline{\mathrm{p}}$.}
    \label{fig:bgonly}
\end{figure}

In Fig.~\ref{fig:bgonly} we report the number of atoms seen by the detector as a function of the number of $\overline{\mathrm{p}}$ in the Cusp trap. All FI are off for this test. One sequence of plasma manipulation routines, optimized for $2.5\times 10^6$ $\overline{\mathrm{p}}$, is used for all experiments. We obtain a linear relation between the number of counts per run and the number of $\overline{\mathrm{p}}$ over a range spanning one order of magnitude. This result suggests that we may obtain a more intense beam by further increasing the number of $\overline{\mathrm{p}}$ used for mixing.

The polarization of the atoms is unknown. On the one hand, an equal distribution among the four hyperfine levels probably overestimates the number of low field seekers because the atoms are formed in a strong magnetic field \cite{robicheaux_2006_threebody}. On the other hand, atoms that decay to $n=1$ within the plasma are not strongly affected since collisions with other $\mathrm{e^+}$ tend to randomize the spin of the bound $\mathrm{e^+}$. For $\upsilon<500\,\mathrm{m/s}$ and $n=1$, trajectory simulations (similar to Ref.~\cite{weiss_2022_simulating}) show that low-field-seeking atoms are far more likely to reach the detector than high-field-seeking atoms. 

Based on our measurements with a hydrogen beam, it was estimated that a hyperfine measurement at $10^{-6}$ precision may be possible with as few as $10^4$ ground state atoms in a weakly polarized $\overline{\mathrm{H}}$ beam \cite{diermaier_2017_inbeam}. The correlation between $n$ and $\upsilon$, demonstrated here for the first time, suggests that the ground state atoms will also be the slowest atoms in the beam. Over $30\%$ of the $\overline{\mathrm{H}}$ in our beam is slow enough for the ASACUSA spectrometer \cite{malbrunot_2019_hydrogen}, and the ground state fraction, roughly $16\%$ of the total, may be completely from that $30\%$. In that case, about 50 slow, ground state atoms are detected in every 15-minute run. The first in-beam measurement of the ground state hyperfine splitting in $\overline{\mathrm{H}}$ may only require 200 such runs or 50 hours of beam time. More time will be needed in reality since the ground state fraction is small (not $100\%$ as assumed in Ref.~\cite{diermaier_2017_inbeam}) and the solid angle of the detector will be reduced by inserting the spectroscopy apparatus in the beamline.

We have also shown that the beam intensity scales linearly with the number of $\overline{\mathrm{p}}$. Whereas presently we use $2.5\times 10^6$ $\overline{\mathrm{p}}$, we can accumulate over $10^7$ $\overline{\mathrm{p}}$ in our catching trap in 10 minutes. If the linear trend in Fig.~\ref{fig:bgonly} continues into the $10^7$ range, then the beam intensity per 15-minute run might be greatly increased through further work to improve $\overline{\mathrm{p}}$ transfer efficiency and parallelization between the trap where $\overline{\mathrm{p}}$ are caught and the trap where $\overline{\mathrm{H}}$ is made. This topic will be explored further in another article focusing on the plasma effects that govern $\overline{\mathrm{H}}$ formation in our experiment.

\section*{Acknowledgments}
We thank Andrew Christensen for sharing his closed-form numerical plasma solver. This work was supported in part by the Istituto Nazionale di Fisica Nucleare (INFN); the Japanese Society for the Promotion of Science (JSPS) KAKENHI Grants No. 19KK0075, No. 20H01930, and No. 20KK0305; the Austrian Science Fund (FWF), Grants No. P 32468 and W1252-N27, the Deutsche Forschungsgemeinschaft (DFG); the National Research Council (Canada); and the Royal Society.

\section*{Data Availability}
The data and analysis that support the findings of this article are openly available \cite{hunter_2025_16352148}.

\bibliographystyle{tfq}
\bibliography{02_bib}

\appendix

\renewcommand{\thefigure}{A\arabic{figure}}
\setcounter{figure}{0}  
\section{Detector data}

\begin{figure}
    \centering
    \includegraphics[width=\linewidth]{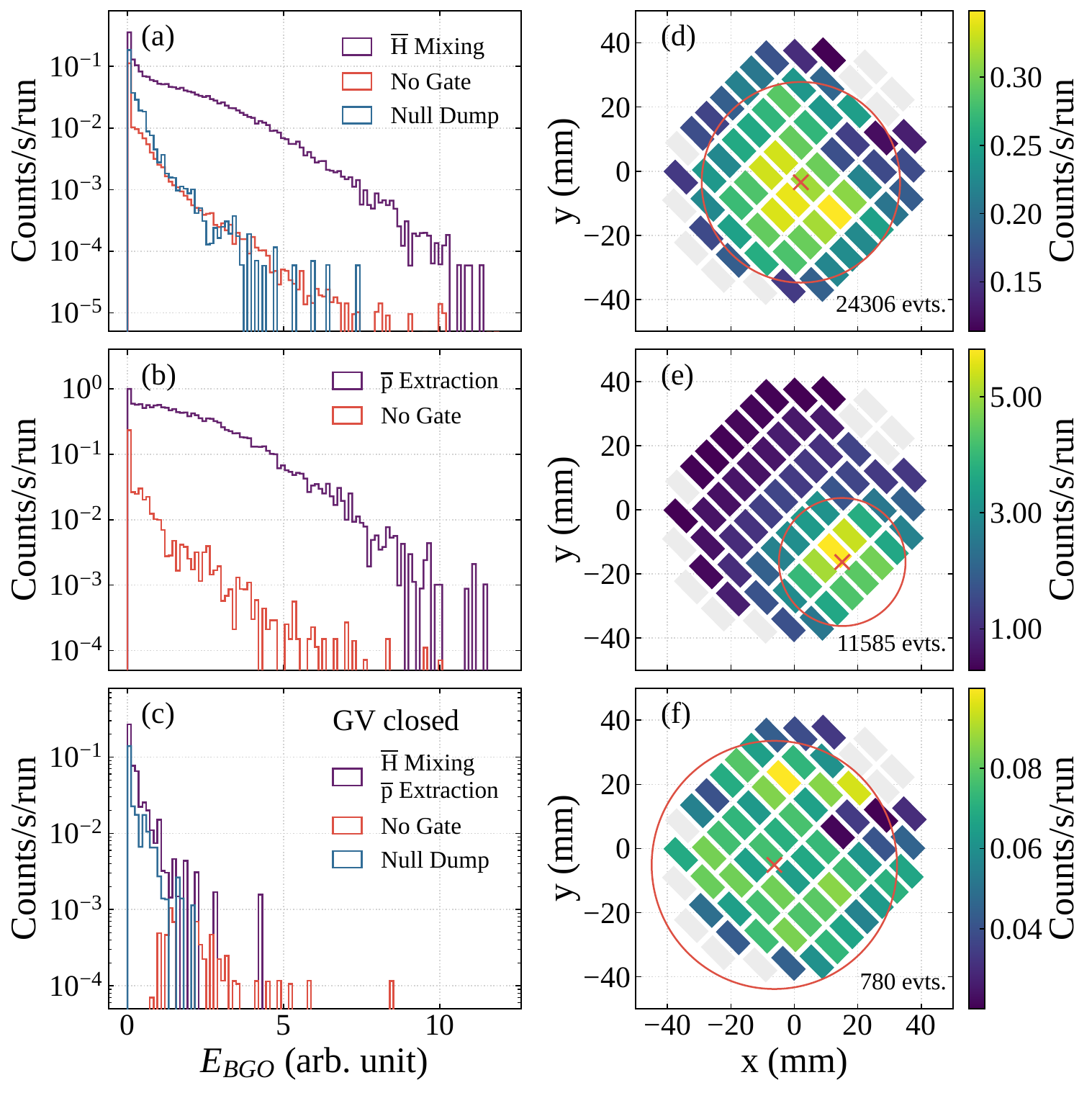}
    \caption{Spectra of energy deposited on the BGO per event (a-c) and hit maps in the transverse plane (d-f). The number of events is given at the bottom right of each plot in (d-f). Channels shown in gray were not working. The red X marks the center of a fitted bi-variate normal distribution and the circle is the 1$\sigma$ contour of a half normal distribution around these centers.}
    \label{fig:hodo_plots}
\end{figure}

Figure~\ref{fig:hodo_plots} summarizes the energy and beam profile data acquired using the detector for 272 $\overline{\mathrm{H}}$ mixing runs (a,d), 18 $\overline{\mathrm{p}}$ extraction runs (b,e), and 11 GV-closed mixing runs plus 4 runs of only cosmic rays (c,f). Panels (a,c) include ``Null Dump" runs where $\overline{\mathrm{p}}$ are extracted downstream with about $1\,\mathrm{eV}$ of kinetic energy. In this case the $\overline{\mathrm{p}}$ annihilate at the field null ($B=0$ at $r=0$) close to the upstream FI. Panels (a,b,c) also include ``No Gate'' data (signals acquired during the whole 15 minute run except the 60 second window(s) for mixing, extraction, and/or null dump). The counts are normalized to acquisition time. The energy in (a-c), in arbitrary units (arb.\ unit), is the sum of the time-over-threshold in $\mathrm{\mu s}$ from all silicon photomultipliers assigned to the BGO. An energy calibration has been performed up to $45\,\mathrm{MeV}$ using signals from cosmic rays. 

The deposited energy is similar for $\overline{\mathrm{p}}$ and $\overline{\mathrm{H}}$ hitting the BGO, only the rate is higher for antiprotons. During runs with the GV closed, events with an energy deposit above $3$ arb.\ unit are rare. The mean energy is 1.41, 1.73, 0.28 arb.\ unit for $\overline{\mathrm{H}}$, $\overline{\mathrm{p}}$, and GV closed. The $\overline{\mathrm{H}}$ beam is significantly broader than the beam of extracted $\overline{\mathrm{p}}$, while GV-closed and cosmic runs have no spatial structure. The centers of the distributions are also different. The center of the $\overline{\mathrm{p}}$ distribution is determined by the residual $B$ outside the trap and can be changed by placing sheets of mu-metal near the detector.



\label{sec:appx_hodo}

\renewcommand{\thefigure}{B\arabic{figure}}
\setcounter{figure}{0}  
\section{Field ionization}

\begin{figure}
    \centering
    \includegraphics[width=\linewidth]{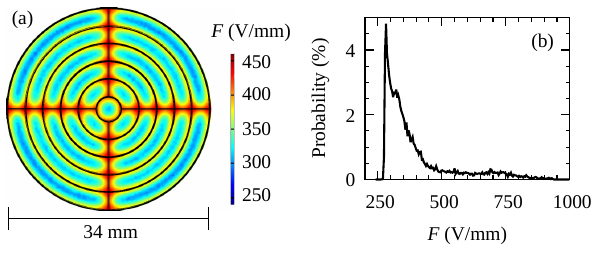}
    \caption{Electric field around the FI inside the trap. (a) Cross section in the midplane between the two grids. Units of color bar are V/mm when $U=1000\,\mathrm{V}$. FI tines are outlined in black. (b) Histogram of the maximum field $F$ seen on all trajectories parallel to the beam axis and through the openings in the FI.}
    \label{fig:comsol}
\end{figure}

We use numerical methods to find the probability that an atom of given principal quantum number $n$ is ionized when passing through one of the FI at a given bias voltage $U$. The FI are pairs of grids with $0.3\,\mathrm{mm}$ thickness and $2\,\mathrm{mm}$ axial separation. We can bias either the downstream side of the upstream FI or the upstream side of the downstream FI up to $\pm 1000\,\mathrm{V}$. Figure~\ref{fig:comsol} shows a rendering of the FI and the electric field when $U=1000\,\mathrm{V}$. For comparison with previous work \cite{kolbinger_2021_measurement}, we also use the EFI, taking the average electric field to be $91\%$ of the bias divided by $10.25\,\mathrm{cm}$. 

\begin{figure}
    \centering
    \includegraphics[width=\linewidth]{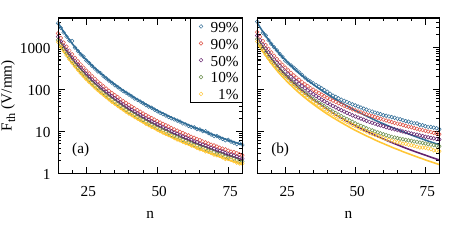}
    \caption{Range of ionization threshold for $B=0$ (a) and $B=0.35\,\mathrm{T}$ (b). The colors correspond to the percentage of atoms ionized after $0.5\,\mathrm{ns}$. The solid curves (yellow, purple, blue) are the $B=0$ prediction \cite{rakovic_1998_ionization}, $F_{th}^{-1}=(7.7,6.0,2.6)\times n^4$ for $(0,50,100)\%$ ionization.}
    \label{fig:Fvn}
\end{figure}

Next, we calculate the probability that an atom with a given $n$ is ionized by a given $F$. We simulate a classical electron orbiting a stationary proton using adaptive Runge-Kutta \cite{garcia_2000_numerical}. We scan the size, ellipticity, and angle of the orbit through values corresponding to $15\leq n\leq 80$ and all possible values of the other quantum numbers. The electron is propagated for $1\,\mathrm{ns}$ or until it reaches $10$ times the nominal orbital radius ($n^2$ times the Bohr radius). We get similar results with longer simulation time, but $1\,\mathrm{ns}$ is sufficient to obtain good agreement with theory at $B=0$ for the range of $n$ studied. $F$ is ramped linearly from $0$ to its nominal value over the first $0.5\,\mathrm{ns}$, which is slow enough that the results are independent of ramp time. We scan the strength and direction of $F$, defining $F_{th}$ as the field value above which the electron is no longer bound after $0.5\,\mathrm{ns}$. The results are given in Fig.~\ref{fig:Fvn} for two values of $B$.


The simulation agrees with the theoretical limits given in Ref.~\cite{rakovic_1998_ionization} for $B=0$. As $B$ increases, it becomes harder to ionize high-$n$ states. $B$ increases with radius at the FI: $0.1<B<0.35\,\mathrm{T}$. 
Rather than repeat the simulations for Fig.~\ref{fig:Fvn} for all $B$ values, we use the $B=0.35\,\mathrm{T}$ simulation to get a conservative estimate for the lowest $n$ ionized at each value of $F$. 

Finally, we obtain the range of $n$ for which atoms are ionized when a given voltage is applied to the FI. We combine the distribution of electric fields in Fig.~\ref{fig:comsol}(b) (or a flat distribution for the EFI) with the probability of ionization vs.\ electric field in Fig.~\ref{fig:Fvn}. We use $0.00\,\mathrm{T}$ for EFI and $0.35\,\mathrm{T}$ for US and DS FI.  

\begin{figure}
    \centering
    \includegraphics[width=0.85\linewidth]{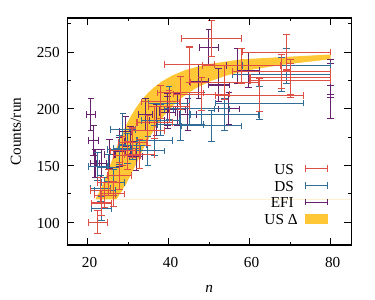}
    \caption{3-coincidence counts at the detector for variable field ionizer bias. Bias is transformed to error bars in $n$ ($10\%$ to $90\%$ ionization range) as in Fig.~\ref{fig:ndist}(b). Y error bars are given as $\sqrt{Y}$, although counting statistics may not be the dominant source of error. FI polarity is negative for US, positive for DS, and positive for EFI (with upstream side of US, downstream side of DS, and upstream side of EFI grounded). We also replot the $B=0.35\,\mathrm{T}$ data from Fig.~\ref{fig:ndist}(b), labeled ``US $\Delta$,'' for which the US FI polarity is positive and the plasma preparation is different (more $\overline{\mathrm{p}}$). The data is scaled by a factor 640 to more easily compare the time-series data with US, DS, EFI (per-cycle data).}
    \label{fig:compare}
\end{figure}

We can test the foregoing analysis using additional experimental data. We apply a given DC voltage $U$ (no pulsing) to one of the FI or to the EFI for one entire run and convert $U$ to $n$ ranges. $U$ is transformed to $10{-}90\%$ ionization ranges in $n$ using the distribution of $F$ (Fig.~\ref{fig:comsol}(b) for the FI, or a single averaged value for the EFI) and the distribution of $n$ given $F$ (Fig.~\ref{fig:Fvn}), with the latter calculated at $B=0.35\,\mathrm{T}$ for the FI and $B=0$ for the EFI. We compare the number of atoms detected as a function of $n$. If our analysis is correct, then the data should roughly overlap. Figure~\ref{fig:compare} shows the results.

All four configurations show a similar dependence of beam intensity (counts) on the threshold $n$ values for field ionization. Most of the ionizable part of the beam seems to be in the range $20<n<50$. The overall agreement validates the analysis in this section. It also implies that in-flight spontaneous decay between the ionizers has little effect on the $n$ distribution. The two high points in the ``EFI'' series must have another cause, since all atoms passing at high field should also pass the EFI at lower field. Using the MCP to detect ions, we observe an increase in ion flux from the EFI at very high bias.
\label{sec:appx_fi}

\renewcommand{\thefigure}{C\arabic{figure}}
\setcounter{figure}{0}  
\section{Time of flight analysis}

\begin{figure}
    \centering
    \includegraphics[width=\linewidth]{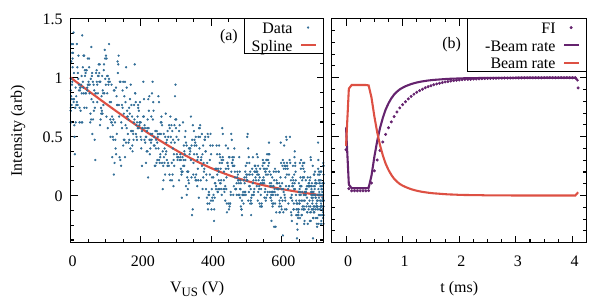}
    \caption{Calculating the time-dependent beam intensity downstream of the FI. (a) Fitting the intensity as a function of voltage, (b) converting the measured time-dependent voltage using (a). The purple curve in (b) is (-1) times the red curve, for easier comparison with the FI voltage.}
    \label{fig:ToF}
\end{figure}

Figure~\ref{fig:ToF} shows how we model the number of ionizable atoms passing the FI as a function of time. We use the data from Fig.~\ref{fig:beam}(c) to find the detector signal as a function of $V_\mathrm{US}$, fit a spline to it, and scale the spline to (0,1) as shown in panel (a). We use the spline as a weight to convert $V_\mathrm{US}(t)$, measured via a $10000{:}1$ resistive divider, into predictions for the relative beam intensity, shown in panel (b). 

For each candidate $T_m$ value, we use the corresponding $f(\upsilon)$ to produce a model ToF curve. Let $d$ be the distance from US FI to detector. Each point on the ``Beam rate'' curve in Fig.~\ref{fig:ToF}(b) gets delayed by the remainder of $(d/\upsilon) / (1/\nu)$, where $\nu$ is the pulse repetition rate, and weighted by $f(\upsilon)$ for every $\upsilon$ in the distribution. The sum of those weighted points is the model curve. The model curve of each $T_m$ is fit to (a bootstrap subsample of) the data using the lm() function in R, and the one with the lowest residual sum of squares is chosen. The final $T_m$ and its uncertainty are given as the mean and standard deviation of the set of $T_m$ obtained for $10^4$ random subsamples.
\label{sec:appx_tof}

\end{document}